\documentclass[a4paper,12pt]{article}
\usepackage{amsmath}
\usepackage{amssymb}
\usepackage[top=3cm,bottom=3cm,left=3cm,right=3cm]{geometry}
\newcommand{\Tr}{\operatorname{Tr}}
\usepackage{xcolor}
\usepackage{amsthm}
\newcommand{\dd}{\,\mathrm{d}}
\theoremstyle{plain}

\begin{document}
\author{\parbox{\textwidth}{\centering
Levent Akant$^{1}$, 
Ebru Do\u{g}an$^{2}$, 
Emine Ertu\u{g}rul$^{3}$ 
\\ O.~Teoman Turgut$^{4}$\\
$^{1,2,3,4}$Department of Physics, Bo\u{g}azi\c{c}i University, Bebek, 34342 Istanbul, Turkey\\[2mm]
$^{1}$akant@bogazici.edu.tr, 
$^{2}$edogan@physics.umass.edu\\
$^{3}$emine.ertugrul@bogazici.edu.tr,
 $^{4}$turgutte@bogazici.edu.tr
}}

\title{\bf On  the  Thermodynamic Limit of  Bogoliubov's  Theory of  the Bose Gas}
\maketitle
\begin{abstract}
Assuming that Bogoliubov's theory of weakly interacting dilute Bose gas defines a self-consistent  model Hamiltonian, we investigate its thermodynamic limit as we take the volume to infinity. The infinite volume is taken via a sequence of  scaled  convex regions with  piecewise smooth boundary and the volumes staying proportional to the cube of the diameter of the region. To get a strict bound on the behavior of the thermodynamic limit, we use the recent formulation of Bogoliubov's theory of condensation  in terms of heat kernels for a given domain as well as an  estimate of the difference of traces between the  heat kernel with Neumann boundary conditions on this domain and the infinite space  result. We cannot control the limiting process by  the area term; however, we can come arbitrarily close to it. 
\end{abstract}

\section{Introduction}

There is considerable interest in understanding how a macroscopic system approaches its thermodynamic limit. One would expect that, as the volume of the system goes to infinity in a precise sense, there should be an order-by-order expansion, the first term of which is the so-called bulk result and the remaining terms are given by lower order terms compared to the volume, typically a series in the length scale of the system lower than the volume term. The common case of this expansion is a term proportional to the area and then the typical length scale for the size of the system and so on.
The thermodynamic limit of quantum systems has been investigated by many researchers. Since we are interested in the Bose-Einstein condensation (BEC), we will review only some of the previous works in this direction. A series of papers by Pathria and collaborators investigated the limit for a free gas when the particles are confined into a rectangular box.  By means of the Poisson summation formula (or adaptations thereof), one can identify the bulk result immediately; then, the  remainder terms, that is the terms that come as corrections to the bulk result are estimated by means of summation techniques (some parts of which they develop)\cite{Pathria0, Pathria1,Pathria2,Pathria3, Pathria3-2, Pathria3-3}. The basic idea is to write the total number of particles for  $N$ bosons confined in a box of sides $L$, under the Neumann boundary conditions. That leads to an expression,
\begin{equation}
N=\sum_{n_1,n_2,n_3} \sum_{j=1}^\infty e^{j\mu \beta}e^{-\pi j ({\frac{\lambda} {L}})^2(n_1^2+n_2^2+n_3^2)}
,\end{equation}
where the usual thermal wavelength is given by $\lambda={h\over \sqrt{2\pi m kT}}$ and $n_1,n_2,n_3$ are integers. Application of the Poisson summation formula to this series leads to
\begin{equation}
N={L^3\over \lambda^3}\sum_{j=1}^\infty {e^{j\mu\beta}\over j^{3/2}}+{L^3\over \lambda^3} \sum_{j=1}^\infty \sum'_{ q_1,q_2,q_3} {e^{j\mu\beta}\over j^{3/2}} e^{-{\pi L^2\over j \lambda^2}( q_1^2+q_2^2+q_3^2)}
,\end{equation}
where the prime indicates that the integers, $q_1, q_2,q_3$,  cannot all be equal to zero.
The first term corresponds to the bulk result, and the next terms correspond to the higher order corrections. As long as $\mu\neq 0$, the last term can be shown to be of a smaller order. When $\mu\to 0^+$ (in fact, to $\approx {1\over L^3}$), the last term contains an undetermined constant of order $L^3$, which can be identified as  condensation, and the remaining sum can be bounded by a term of the form $L^2/\lambda^2$. Pathria et al. have shown by a careful analysis that the sum has a leading behavior given by,
\begin{equation}
N={L^3\over \lambda^3}\sum_{j=1}^\infty {e^{j\mu\beta}\over j^{3/2}}+N_0+C {L^2\over \lambda^2}
,\end{equation}
where the constant $C$ is determined through a convergent sum.
 Hence, the limit $L\to \infty$ can be taken to yield the thermodynamic limit. Relativistic extensions of this approach are also considered in \cite{Pathria4, Pathria5}. This is a simple and intuitive picture of condensation in the thermodynamic limit.

The interacting Bose gas is naturally a more complicated problem. One of the first  attempts is in \cite{Angelescu-Nenciu}, where it is shown that  the thermodynamic limit is independent of the boundary conditions for bosons interacting with well-behaved two-body potentials by means of heat kernel comparisons. They derive some useful bounds for Dirichlet heat kernels on a convex domain $\Omega$ of the kind,
\begin{equation}
\Big|{\rm Tr}(e^{t\Delta_D})- {V(\Omega)\over (4\pi t)^{d/2}}\Big|\leq {e^{d/2} V(\partial\Omega)\over 2(4\pi t)^{(d-1)/2}},
\end{equation}
 where $-\Delta_D$ refers to the Dirichlet Laplacian; these bounds are essential to understand the thermodynamic limit of bosonic systems.
They obtain similar estimates for general boundary conditions and as a result obtain the independence of the thermodynamic limit from boundary conditions.
 (The inequality analogous to this for the difference of the Neumann heat kernel of the domain and the heat kernel for the unbounded space presented in this work, contains a factor $e^{\lambda t}$, hence cannot be used for our purposes). 
The rigorous studies of BEC for a free Bose gas are given by Lewis and Pule \cite{Lewis} as well as by  Landau
and Wilde \cite{landau}. The work of van den Berg gives an elegant and appealing picture of the thermodynamic limit of the free Bose gas \cite{van den Berg1}. A rigorous derivation of the noninteracting BEC taking into account effects of different shapes of containers, hence leading to possible macroscopic occupation of excited levels,  is given by van den Berg, Lewis and Pule  in \cite{vandenberg-lewis, vandenberg-Pule2}, which opened up a critical analysis of what is meant by condensation.
 Subsequently, van den Berg gave \cite{van den Berg2} an improved estimate of the heat kernel for convex domains with boundaries having  positive curvature bounded from above by ${1\over R}$ ($R>0$). As a result, he  showed that the partition function for a Boltzmann gas has area term corrections rigorously bounded by the area of the boundary times the curvature of the boundary, moreover, his result shows that  one can also get a higher order  bound, if the curvature also goes to zero by the size of the system, that is if $R \propto  V(\Omega)^{\alpha}$ (for $\alpha >0$). Consequently, he gave an estimate,
\begin{equation}
\left|{\rm Tr}(e^{t\Delta_D})- {V(\Omega)\over (4\pi t)^{d/2}}+{V(\partial \Omega)\over 4 (4\pi t)^{(d-1)/2}}\right|\leq {V(\partial \Omega)\over (4\pi t)^{(d-2)/2} R}(c_1(d)+c_2(d)\ln[1+{R^2\over t}]),
\end{equation}
where $c_1(d), c_2(d)$ are  explicitly known positive constants dependent on dimension. This is an interesting expression to have if one wants to control the area corrections related to the thermodynamic limit. Again, they are also important for understanding BEC. Hence, if the curvature also goes to zero sufficiently fast as the domain size grows,  these correction terms are negligible.  Further work on the Dirichlet Laplacian along similar lines can be  found in  \cite{van den Berg3,van den Berg4}.  Our approach here  is  very much inspired by these previous  works.
 These results suggest that indeed the approach to the thermodynamic limit can  be organized  as a bulk term and a term proportional to the area of the box and then following lower order terms for the usual containers where area is of lower order. Here, one assumes that the curvature terms due to boundary  get smaller as the size of the confining  region grows. This idea can indeed  be worked out
for a general convex body with smooth boundary,  by means of an asymptotic expansion of the heat kernel. In their work, Kirsten and Toms, using short time heat kernel asymptotics and Mellin-transform representation of the exponential function, gave an asymptotic expansion of various thermodynamic variables for a free Bose gas \cite{toms-kirsten}.  
Nevertheless, this does not give a true control on the remainder terms as one takes  the thermodynamic limit, so it is not  suitable for the present purpose.

For interacting particles, there is a considerable amount of rigorous work by Lieb, Yngvason, Seiringer. (see \cite{Liebbook} and references therein). These works are concerned with obtaining rigorous bounds on various macroscopic quantities such as the ground state energy of the system, pressure of the condensate. Moreover, they  investigate rigorously under what conditions the weakly interacting gas description would be consistent. A landmark contribution by E. Lieb and R. Seiringer shows rigorously that  trapped dilute Bose gas indeed exhibits condensation into a state which minimizes the Gross--Pitaevskii functional\cite{Lieb-seiringer}. It is essential from a conceptual point of view to justify condensation starting directly from many body Hamiltonian, since Bogoliubov theory takes as an assumption the existence of the condensation. A considerable amount of work is done by Erdos et al. \cite{erdos1, erdos2, erdos3, erdos4} to construct a rigorous derivation of Gross--Pitaevski equation for the evolution of the ground state starting from the basic microscopic model. To rigorously study such systems these works typically  assume that the system is in a rectangular box; moreover the results are not usually organized as the bulk result plus a uniform next order correction term,  such as an area contribution.  A comprehensive review of weakly interacting Bose gas, with precise statements, is given in the excellent review by  Zagrebnov and Bru \cite{Zagrebnov}. It is known that it is not possible to improve the Bogoliubov approximation while retaining self-consistency; therefore, one needs to go beyond this approximation. Nevertheless, from a rigorous point of view, Bogoliubov theory \emph{per se} seems to be  not a completely  consistent description of the weakly interacting gas either \cite{Zagrebnov} (apart from the well known problem of calculating the critical temperature from this theory).  For the current state of affairs, we also refer the reader to the article by R. Seiringer \cite{seiringer-JMP}. Recently, more refined proofs for the Bogoliubov excitation spectrum in the thermodynamic limit have been established for bosons in a box and for trapped Bose gases\cite{Boccato2019, Nam2021}, providing further justification for the model's validity.

Yet, it would be very appealing to have a more intuitive and simpler picture of the thermodynamic limit in the interacting Bose gas, even if one assumes a priori that this weakly interacting system is well approximated by the Bogoliubov approach. Moreover, assuming that the weakly interacting Bose gas has  a consistent description in the Bogoliubov picture, it would be very informative to see how the thermodynamic limit is achieved for such a model. This would be along the lines of previous works by Pathria et al. and van den Berg et al. for this particular model.
In the present work, we assume that we have a model of the weakly interacting Bose gas in its condensing phase described by the Bogoliubov Hamiltonian in a finite but large box, which is convex and in general has a piecewise smooth boundary (this is to allow for edges only, but we have in general smooth boundaries in mind; nevertheless, the heat kernel  estimates used are true for Lipschitz boundaries) as is developed in our previous work \cite{akant1}. In this model,  we investigate the thermodynamic limit, and show that indeed the bulk result is achieved by error  terms of order arbitrarily close from above (by a factor of an arbitrarily small power of the length scale)  to the area of the domain; however, we will not be able to show that it is exactly equal to the area terms due to estimates that we use for the Neumann heat kernel of a  general domain. It is possible that this is due to the simplicity of our approach, and a more sophisticated method may remove this defect.

In our approach it was most convenient to use the Neumann boundary conditions for the eigenvalues of the Laplacian on a (convex) domain; it remains as a technical challenge to develop a similar model for the Dirichlet boundary conditions, where many powerful estimates of heat kernels do exist.
 As shown previously in our work, the ground state energy of the weakly interacting gas, as found by Lee and Yang, follows  from our expression for a general domain, when we replace the heat kernel with the flat space expression. It is quite remarkable that a rigorous justification of this second order correction has been achieved recently by H-T Yau and J. Yin\cite{yau-lee-yang}.  Building upon these results, the second-order correction to the ground state energy has seen further significant rigorous developments, providing a more complete understanding of the dilute Bose gas energy beyond the initial Bogoliubov approximation \cite{Fournais2020}. Similarly, the depletion coefficient also goes over to the usual bulk result, when the heat kernel is replaced directly with the usual  unbounded Euclidean  space result.

The advantage of our approach is to formulate all the relevant thermodynamic expressions  in terms of the heat kernel of the Laplacian on this domain. Therefore, if one can find good global estimates of the heat kernel in a given domain, this will provide estimates of thermodynamic variables of the system. Let us emphasize that this does not mean  a construction for the thermodynamic limit of the  weakly interacting Bose systems, that would require one to start from the original many body Hamiltonian, establish the condensation and validity of the assumptions of Bogoliubov theory while taking the thermodynamic limit in a manner controlled by, preferably, the  area term. 

 In the present work, we  show that, in  the thermodynamic limit, all (physically) relevant quantities  go to the flat bulk results, assuming the Bogoliubov model of weakly interacting Bose gas.  We assume that we have a nested sequence of convex regions; one can think of it as scaled versions of an initial large body. For such a convex body, which we denote as $\Omega$, we assume that the volume goes  as $D_\Omega^3$  where $D_\Omega$ refers to the diameter of the region.  We keep the assumed condensation density $n_0$ constant, while the volume is sent to infinity. To achieve our goal of precisely controlling the thermodynamic limit,  it is crucial to have an estimate of the Neumann heat kernels. Although there are doubts about the complete consistency of the Bogoliubov model, we believe that understanding the thermodynamic limit within the realm of this theory brings some valuable insight into the theory of weakly interacting dilute  systems.

\section{Ground State Energy}

Let us  state the formula for the energy that is found in equation (106) of \cite{akant1} (within the approximations of Bogoliubov theory); here, we take  $a=u_0n_0$, which is assumed to be a  small parameter and {\it kept constant} during the thermodynamic limit. We recall that the $c$-number substitution used is an effect of order $\ln (V)/V$, thus negligible in the thermodynamic limit; moreover, it is of {\it lower order} than the surface type terms that we will obtain; hence it is consistent to continue within this approach. Ignoring higher order interactions may not be small in this sense but it is assumed to be negligible in the Bogoliubov approach. Based on these, we solely focus on the Bogoliubov Hamiltonian as a model system. As a result we have the ground state energy of the system,
\begin{equation}\label{GSnrg}
  E_{gr}=\frac{u_{0}n_{0}^{2}V}{2}+\frac{a}{2\pi}\int_{0}^{\infty}\dd t\int_{0}^{1} \dd y\,F(t,y){\rm Tr}'e^{\Delta t/a}
\end{equation}
where
\begin{equation}
  F(t,y)=\sqrt{1-y^{2}}(1-e^{-t(1-y)}+1-e^{-t(1+y)}).
\end{equation}
Using the simple identity
\begin{equation}\label{replacement}
(1-e^{-a(1\pm y)t})=\int\limits_{0}^{1} \dd\zeta \,\, e^{-a(1 \pm y)t\zeta} \,\, (a(1 \pm y)t),
\end{equation}
we get
\begin{eqnarray}
&&E_{gr}=\frac12\,u_0 n_0^{2} V \nonumber\\
&&+\frac{a^{3}}{2\pi}\int_{0}^{\infty} \dd t\int_{0}^{1} \dd y\int_{0}^{1} \dd \zeta  \,t\,\sqrt{1-y^{2}}\times \Big[(1-y)e^{-a(1-y)t\zeta}+(1+y)e^{-a(1+y)t\zeta}\Big]\nonumber\\
&& \int_{\Omega} \dd X\, \Big[ K_t(X,X)\,-\frac1V\Big]
\label{eq:Egr}
\end{eqnarray}
where $dX$ refers to the integration over the domain $\Omega$ in ${\mathbf R}^3$.

Central to our work, we use a remarkable estimate given in a paper by Brown on Lipschitz domains with Neumann boundary conditions \cite{Brown}. There we keep $0<\eta<1$ in the general expression, which has an estimate different from the
Dirichlet boundary conditions by a prefactor $\Big[ \partial (X)\Big/ t^{1/2}\Big]^\eta$, and the coefficient in front is given as $C_\eta$ (which, in principle,  may grow as we let $\eta\to 0^+$)
\begin{equation}\label{brown}
\left|K_{t}(X,X)-\frac{1}{(4\pi t)^{3/2}}\right| \leq \underbrace{C_{\eta} \left(\frac{\partial(X)}{\sqrt{t}}\right)^{\eta}\,\frac{e^{-\partial^{2}(X)/C\,t}}{(4\pi t)^{3/2}}}_{(*)}
\end{equation}
Here for any point $X$ in the domain $\Omega$,  $\partial(X)$ refers to the distance to the boundary surface $\partial \Omega$ of $\Omega$ and $C$ is a sufficiently large constant which depends only on the domain $\Omega$. To the best of our understanding, to have uniform control over the domain $\Omega$,  one needs this power factor controlled by some $\eta>0$, which unfortunately has no natural explanation in Bogoliubov theory.

In two special cases the volume integration $dX$ can be easily written as a product measure.
Using $z=\partial(X)$, for a cube of side length  $L$, we have   $f(z)=6\,(L^2-4Lz+4z^2)$  and $f(z)=4\pi(L-z)^2$ for a sphere of radius $L$. Thus, integral over $dX=f(\partial(X))d\partial(X)$ becomes $f(z)dz$ in these special cases.
In general, the coordinate transformation to $\partial(X)$ and a (usual area)  integration over the corresponding level surfaces may not be possible. However, it is possible to find an upper bound for the integration over the domain as we will discuss shortly.

We replace the heat kernel in the energy expression with the flat formula and then take the difference with our expression, moreover we divide everything by the volume $V(\Omega)$ to eventually take the thermodynamic limit. This difference can be bounded from above by replacing the heat kernel in the energy expression with the estimate $(*)$ given above, since all other terms are positive functions. This is what we will use for our estimates, since  in general the formulae we have all contain positive terms as we will see.

Since in our energy formula the potentially more complicated part is $1-e^{-(1-y)at}$, we put aside the other term for the time being and focus on estimating this combination only which we call for now  as $(**)_1 $. The other part, which we name as $(**)_2$ and  involving the term $1-e^{-(1+y)at}$  actually behaves better, and can be dealt with similar techniques.
The part to be estimated is the right-hand side convoluted with the expressions that we have in  the ground state energy formula, which we denote by $(**)_1$. For simplicity we denote $\partial (X)$ by $z$ in our formulae and keep the volume as $dX$ till the end.
\begin{align}
(**)_1 &\leq a^2\int_\Omega {\dd X\over V(\Omega)} \int_0^\infty \dd t\int_0^1 \dd y\int_0^1 \dd \zeta \,\, \,\,C_{\eta} \left(\frac{z}{\sqrt{t}}\right)^{\eta}\, \frac{e^{-\frac{z^2}{C\,t}-a(1-y)t\zeta}}{t^{3/2}}\,a(1-y)t \, (\sqrt{1-y^2})  \nonumber\\
&\leq a^3C_{\eta} \int_\Omega {\dd X\over V(\Omega)} \int_0^\infty  \dd t\int_0^1 \dd y\int_0^1 \dd \zeta \,\,\left(\frac{z}{\sqrt{t}}\right)^{\eta}\, \frac{e^{-\frac{z^2}{C\,t}-a(1-y)t\zeta}}{t^{1/2}} \, (\sqrt{1+y})(1-y)^{3/2}\nonumber
\end{align}
% --- WARNING: constants absorbed into C_\eta ---
\noindent\textbf{Warning.}
From this point on, to keep the expressions simpler, we will repeatedly \emph{absorb} various numerical constants
coming from crude inequalities (e.g.\ $\sqrt{1+y}\le\sqrt{2}$), changes of variables,
and various uniform bounds, that we use,  into the coefficient $C_\eta$. Thus the actual numerical value of $C_\eta$ may change from line to line.
% ----------------------------------------------

If we scale out the variable $z$ in $t$ and use  $\sqrt{1+y}<\sqrt{2}$, after absorbing various constants into $C_{\eta}$ again,
and making the change of variable in the $t$ integral as $u=1/t$, we get,

\begin{eqnarray}
(**)_1&\leq& a^3 C_{\eta}\int_\Omega {\dd X\over V(\Omega)} \int_0^1 \dd y\int_0^1  \dd \zeta \,z (1-y)^{3/2} \int_{0}^{\infty} e^{-u-\frac{a(1-y)\zeta z^2}{C\,u}}\, \frac{\dd u}{u^{1+\frac{1}{2}-\frac{\eta}{2}}}\nonumber\\
&\leq& a^3C_{\eta}\int_\Omega  {\dd X \over V(\Omega)}\int_0^1  \dd y \int_0^1{\dd\zeta\,z^{1/2+\eta/2}}{a^{-1/4+\eta/4} (1-y)^{3/2-1/4+\eta/4} \zeta^{-1/4+\eta/4}}\nonumber\\
&\ & \quad \quad \quad \quad \quad \quad \quad \quad \quad \times K_{1/2-\eta/2}\!\Big(2z\sqrt{\tfrac{a(1-y)\zeta}{C}}\Big)
\nonumber
\end{eqnarray}
We now use the following integral representation of the Bessel function,
\begin{equation}
K_{\nu}(x) = \frac{\pi}{\sqrt{2 x}} \frac{e^{-x}}{\Gamma(\nu+\frac{1}{2})} \int_{0}^{\infty}\dd s\, e^{-s} s^{\nu-1/2} {\left(1 +\frac{s}{2x}\right)}^{\nu-1/2}
\end{equation}
Hence,
\begin{eqnarray}
&&(**)_1\leq 
a^{5/2+\eta/4} C_{\eta}\int_\Omega {\dd X\over V(\Omega)} \int_0^1 \dd y \int_0^1{\dd \zeta \,z^{\eta/2}} (1-y)^{1+\eta/4} \zeta^{-1/2+\eta/4}\nonumber\\
&\ & \ \ \ \ \ \ \ \ \ \ \ \ \ \ \  \ \times e^{-2z\sqrt{\frac{a(1-y)\zeta}{C}}}\int_0^\infty
\dd s\, e^{-s} s^{-\eta/2}\Big( 1+\frac{s}{4z\sqrt{\frac{a(1-y)\zeta}{C}}}\Big)^{-\eta/2}
\end{eqnarray}
Note that the last term of the last integral is smaller than $1$ since $\eta>0$, that is,
\begin{equation}
\Big( 1+\frac{s\sqrt{C}}{2z\sqrt{a(1-y)\zeta}}\Big)^{-\eta/2}<1,
\end{equation}
 there is no convergence issue since $\eta<1$; thus, simplify the integral estimate,
\begin{eqnarray}
(**)_1&\leq& a^{5/2+\eta/4} C_{\eta}\int {\dd X\over V(\Omega)} \,z^{\eta/2} \int_0^1 \dd y \, (1-y)^{1+\eta/4}\int_0^1 \dd \zeta \zeta^{-1/2+\eta/4}  e^{-z\sqrt{\frac{a(1-y)\zeta}{C}}}\nonumber
\end{eqnarray}
If we make a change of variable $\zeta=\xi^2$ we can estimate the integral as,
\begin{equation}
\int_0^1 \dd\zeta \zeta^{-1/2+\eta/4}  e^{-2z\sqrt{\frac{a(1-y)\zeta}{C}}}=\int_0^1 \dd\xi \xi^{\eta/2} e^{-2z\xi\sqrt{\frac{a(1-y)}{C}}}\leq \int_0^1 \dd\xi  e^{-2z\xi\sqrt{\frac{a(1-y)}{C}}}< \frac{\sqrt{C}}{2z\sqrt{a(1-y)}}\nonumber
\end{equation}
Thus we find,
\begin{eqnarray}
(**)_1&\leq& a^{5/2-1/2+\eta/4} C_{\eta}\int_\Omega {\dd X\over V(\Omega)} \,z^{\eta/2-1} \int_0^1 \dd y \, (1-y)^{1/2+\eta/4}\leq  a^{2+\eta/4} C_{\eta}\int \dd X \,z^{\eta/2-1} \nonumber
\end{eqnarray}
If we take the special case of cubic box of sides $L$, the measure can be decomposed as $dX=6\,(L-2z)^2dz$,
hence,
\begin{equation}
(**)_1< a^{2+\eta/4} C_\eta {1\over L^3} \frac{L^3}{L^{1-\eta/2}}=a^{2+\eta/4} C_\eta \frac{1}{L^{1-\eta/2}}
\end{equation}
As remarked above, once we work out the other term, a very similar calculation produces a very similar bound, hence as a result we get control over the thermodynamic limit.

\paragraph{A geometric bound for distance integrals.} 
Here we make a short digression and 
 state the following observations for any convex domain $\Omega$.  Let us consider the integral of a {\it non-increasing} (positive and  continuous inside the domain) function $g$ of distance $\partial (x)$ to the surface $\partial \Omega$, taken over the whole convex body.
 Here we assume that the boundary is a piecewise smooth submanifold (a regular surface)  of the Euclidean space. Regions defined by the distance $\partial (X)$ to the boundary bigger than or equal to some fixed value $z$ are all convex regions \cite{Gromov}. Let us denote their volume by $H(z)$, where $z$ denotes this lower limit  of the  distance from the boundary surface. This function is a {\it monotone decreasing function} of $z$ thanks to convexity. In general one cannot make a smooth transformation to $z$ since $H(z)$ may be non-differentiable beyond a certain value of $z$, due to  possible jumps, or collapsing to lower dimensional regions. However we may define a Riemann-Stieltjes integral of a function $g(z)$ of $z$ over the whole body using these regions,
\begin{equation}
      \int_\Omega  \dd X g(z)= \int_0^{D_\Omega} g(z) \dd H(z)
,\end{equation}
where $D_\Omega$ is the diameter of the region; in general, the level surfaces are smooth up to a distance at which somewhere the normal curves possibly intersect. A better description perhaps is to think about the co-area formula and write down the integration formula accordingly. 
We can now estimate that $|\int \dd H(z) g(z)|<A(\partial \Omega) \int g(z) \dd z$, since the area of the boundaries of these inner regions are always less than the area of $\partial \Omega$ (thanks to the convexity) and $g(z)$ is assumed non-increasing along $z$. Here we use $A(\partial\Omega)$ for the volume of the boundary, since this is the usual area term in three dimensions. Therefore we find the inequality,
\begin{equation}
      \int_\Omega  \dd X g(\partial(X)) < A(\partial \Omega) \int_0^{D_\Omega} \,\dd z\, g(z) \label{geo}
.\end{equation} 
This result  can now be used for our case discussed above, and it  then  implies that, 
\begin{equation}
      \int \dd X \frac{1}{ \partial(X)^{1-\eta/2}} < A(\partial \Omega) \int_0^{D_\Omega} \frac{\dd z}{z^{1-\eta/2}}
,\end{equation}
which gives the general inequality,
\begin{equation}
 (**) < a^{2+\eta/4} C_\eta A(\partial \Omega)D_\Omega^{\eta/2}.
\end{equation}
This shows that now we can take the thermodynamic limit and the approach to the bulk result is controlled by the term on the right side of this inequality for any small but nonzero choice of  $\eta$.

From now on, we often refer to inequality  in (\ref{geo}) to estimate various terms we encounter in our thermodynamic limit.

\section{Depletion Coefficient}

\subsection*{Zero Temperature}

Let us now recall the expression for the $T=0$ depletion coefficient as found in our previous work, the formula (125) of \cite{akant1}:
\begin{equation}
   n_e(0)=\frac{a}{2}\int_{0}^{\infty} \dd t \int_\Omega \frac{\dd X}{V(\Omega)} K_t(X,X)  e^{-at}I_{1}(at).
\end{equation}
Using the identity
\begin{equation}
\frac{2}{\pi}\int_0^1 \dd y \sqrt{1-y^2} \cosh( uy)=\frac{I_1(u)}{u},
\end{equation}
we can express $ n_e(0)$ as
\begin{equation}
 n_e(0)=\frac{a}{\pi}\int_\Omega \frac{\dd X}{V(\Omega)} K_t(X,X) \int_0^1 \dd y \int_0^\infty \dd t \sqrt{1-y^2} (at) \cosh(aty) e^{-at}
\end{equation}
Let us take the difference $\Delta n_e(0)$ from the infinite space value; we need to bound this difference. For this we use the result given by Brown, again for some $\eta$ with $0<\eta<1$, which, as before, means in our expression for the depletion we replace the heat kernel with $(*)$ in the estimate formula (\ref{brown}),
\begin{equation}
|\Delta n_e(0)|< C_\eta \frac{a}{\pi}\int_\Omega \frac{\dd X}{V(\Omega)}  \int_0^1 \dd y \int_0^\infty \dd t \left(\frac{z}{\sqrt{t}}\right)^{\eta}\,\frac{e^{-z^{2}/(C\,t)}}{(4\pi t)^{3/2}} \sqrt{1-y^2} (at) \cosh(aty) e^{-at}
\end{equation}
This is indeed very similar to our previous estimate; we can write it as
\begin{equation}
|\Delta n_e(0)|< C_\eta \frac{a^2}{\pi}\int_\Omega \frac{\dd X}{V(\Omega)}  \int_0^1 \dd y \int_0^\infty \dd t\, z^{\eta}\,\frac{e^{-z^{2}/(C\,t)}}{ t^{1/2+\eta/2}} \sqrt{1-y^2}[ e^{-at(1-y)} +e^{-at(1+y)}]
\end{equation}
again keeping the term coming  from the first exponential, the other term can be replaced by the same expression since $e^{-at(1+y)}<e^{-at(1-y)}$,  replacing $\sqrt{1+y}$ by $\sqrt{2}$,
\begin{equation}
|\Delta n_e(0)|< C_\eta {a^{1+3/4+\eta/4}}\int \frac{\dd X}{V(\Omega)}  z^{1/2+\eta/2} \int_0^1 \dd y\,  (1-y)^{1/4+\eta/4}K_{1/2-\eta/2}\!\left(2z\sqrt{\frac{a(1-y)}{C}}\right)
\end{equation}
Using very similar arguments, this can be reduced to
\begin{eqnarray}
|\Delta n_e(0)|&<& C_\eta {a^{1+1/2+\eta/4}}\int_\Omega \frac{\dd X}{V(\Omega)}  z^{\eta/2} \int_0^1 \dd y  (1-y)^{\eta/4}e^{-z\sqrt{\frac{a(1-y)}{C}}} \nonumber\\
&<& C_\eta a^{1+\eta/4} \int_\Omega \frac{\dd X}{V(\Omega)}  z^{-1+\eta/2}< C_\eta a^{1+\eta/4}\frac{A(\partial \Omega )}{V(\Omega)}\frac{D_\Omega^{\eta/2}}{\eta}
.\end{eqnarray}
Here we can choose $\eta$ as small as we wish while being positive, then we get the  thermodynamic limit as claimed (note that the convergence gets worse and worse as we lower $\eta$.) As a remark, we remind the reader that we deal with essentially a scaled convex body, hence $A(\partial \Omega)\sim D_\Omega^2$ and volume $V(\Omega)\sim D_\Omega^3$.

\subsection*{Finite--Temperature}

Next, we will establish the same result for the  finite--temperature depletion coefficient, the finite--temperature part of which is given by the formula (142) of \cite{akant1},
\begin{eqnarray}
\tilde n_e(T)
&=&\sum_{k=1}^\infty \Bigg[
\frac{1}{V(\Omega)} \Tr\!\left(e^{-k\beta\Delta}\right)e^{-k\beta a}\nonumber \\
&+& a\int_0^\infty \dd t \;
\frac{1}{V(\Omega)}\Tr\!\left(e^{-\Delta\sqrt{(k\beta)^2+t^2}}\right)
e^{-a\sqrt{(k\beta)^2+t^2}} I_{1}(at)
\Bigg].
\end{eqnarray}
\subsubsection*{First Term of the Finite--Temperature Depletion Coefficient}
We focus on the first contribution
\begin{equation}
\tilde n_e^{(1)}(T)
:=
\sum_{k=1}^\infty
\frac{1}{V(\Omega)} \Tr\!\left(e^{-k\beta\Delta}\right)e^{-k\beta a}.
\end{equation}
Subtracting the free heat-kernel term and using the heat-kernel remainder estimate, this expression is controlled by the right-hand side of the following equation:
\begin{equation}
\frac{1}{V(\Omega)}
\sum_{k=1}^\infty
\left|
\Tr\!\left(e^{-k\beta\Delta}\right)e^{-k\beta a}
-
\frac{e^{-k\beta a}}{(4\pi k\beta)^{3/2}}
\right| \le
C_\eta
\sum_{k=1}^\infty
\int_{\Omega}\frac{\dd X}{V(\Omega)}\;
z^{\eta}
\frac{e^{-z^{2}/(C\,k\beta)}\,e^{-k\beta a}}
{(k\beta)^{3/2+\eta/2}}.
\label{eq:remainder}
\end{equation}
Note that in the $k$-summation we may replace $k^{3/2+\eta/2}$ by $k^{3/2+\eta/4}$, since $k\geq 1$, which yields an upper bound. Define
\begin{equation}
f(k):=\beta^{-\eta/4}\int_{\Omega}\frac{\dd X}{V(\Omega)}\;
z^{\eta}\,
\frac{e^{-z^{2}/(C\,k\beta)}\,e^{-k\beta a}}
{(k\beta)^{3/2+\eta/4}}.
\end{equation}
Then the Euler--Maclaurin formula gives
\begin{equation}
\sum_{k=1}^{\infty} f(k)
=
\int_{1}^{\infty} f(k)\,\dd k
+
\frac{f(\infty)-f(1)}{2}
+
\int_{1}^{\infty} f'(k)
\left(k-\lfloor k\rfloor-\frac12\right)\dd k .
\end{equation}
We now estimate the three Euler–Maclaurin contributions separately.
\paragraph{First Euler-Maclaurin term.}
We now isolate the \emph{integral contribution}
\begin{equation}
\mathcal{I}
:=
\int_{1}^{\infty} f(k)\,\dd k
=
\beta^{-\eta/4}\int_{1}^{\infty}\dd k
\int_{\Omega}\frac{\dd X}{V(\Omega)}\;
z^{\eta}
\frac{e^{-z^{2}/(C\,k\beta)}\,e^{-k\beta a}}
{(k\beta)^{3/2+\eta/4}}.
\end{equation}
Since the integrand is non-negative, we may extend the lower limit \(1\to0\) in \(\mathcal{I}\).
\begin{equation}
\mathcal{I}
\le
\beta^{-\eta/4}\int_{0}^{\infty}\dd k
\int_{\Omega}\frac{\dd X}{V(\Omega)}\;
z^{\eta}
\frac{e^{-z^{2}/(C\,k\beta)}\,e^{-k\beta a}}
{(k\beta)^{3/2+\eta/4}}.
\end{equation}
Setting $t=k\beta a$ we obtain;
\begin{equation}
\mathcal{I}
\le
\beta^{-\eta/4}\int_{\Omega}\frac{\dd X}{V(\Omega)}\;
z^{\eta}
\int_{0}^{\infty}\frac{\dd t}{\beta a}\,
\frac{e^{\left(-\frac{z^{2}a}{C\,t}-t\right)}}
{t^{3/2+\eta/4}}.
\end{equation}
Applying Bessel transformations analogous to those used in the previous sections, we obtain
\begin{equation}
\mathcal{I}
=
\frac{C_\eta}{V(\Omega)\beta^{1+\eta/4}}
\int_{\Omega}\dd X\;
z^{3\eta/4-1/2}
K_{\frac12+\frac{\eta}{4}}\!\big(2z\sqrt{\frac{a}{C}}\big).
\end{equation}
Employing the cosine integral representation of 
$K_\nu(x)$,
\begin{equation*}
  K_\nu(w)={\Gamma(\nu+{1\over 2})\over w^\nu \Gamma({1\over 2})} \int_0^\infty dt {\cos(w t)\over (1+t^2)^{\nu+1/2}} \nonumber 
\end{equation*}
and the inequality $\cos(xt) \leq 1$, we find that the remaining integral is finite, with all constants absorbed into $C_\eta$. Hence,
\begin{equation}
\mathcal{I}
\le
\frac{C_\eta}{V(\Omega)\beta^{1+\eta/4}}
\int_{\Omega}\dd X\; z^{\eta/2-1}.
\end{equation}
Passing to boundary normal coordinates (and using the integration over the level sets), and employing  the fact that the boundary area is largest (thanks to (\ref{geo})),
\begin{equation}
\int_{\Omega}\dd X\; z^{\eta/2-1}
\le
A(\partial\Omega)\int_{0}^{D_\Omega} z^{\eta/2-1}\,\dd z,
\end{equation}
which gives
\begin{equation}
\mathcal{I}
\le
\frac{C_\eta}{\beta^{\eta/4+1}}\,\frac{A(\partial \Omega )}{V(\Omega)}\,D_\Omega^{\eta/2}.
\end{equation}
\paragraph{Second Euler--Maclaurin term.}
We now consider the Euler--Maclaurin contribution coming from the second term
associated with the first contribution, namely
\begin{equation}
\frac{f(\infty)-f(1)}{2}.
\end{equation}
From the definition of $f(k)$ we use the pointwise bound
\begin{equation}
f(k)
\le
C_\eta\, \beta^{-\eta/4}
\int_{\Omega}\frac{\dd X}{V(\Omega)}\;
z^{\eta}
\frac{e^{-z^{2}/(C\,k\beta)}\,e^{-k\beta a}}
{(k\beta)^{3/2+\eta/4}}.
\end{equation}
Since $f(k)\ge0$ and the factor $e^{-k\beta a}$ guarantees decay, we have $f(\infty)=0.$
Moreover,
\begin{equation}
f(1)
\le
C_\eta
\int_{\Omega}\frac{\dd X}{V(\Omega)}\;
z^{\eta}
\frac{e^{-z^{2}/(C\,\beta)}\,e^{-\beta a}}
{\beta^{3/2+\eta/2}}.
\end{equation}
Using $e^{-\beta a}<1$ and passing to the boundary normal coordinates, and then using
 $z^{\eta}\le D_\Omega^{\eta}$ for $0\le z\le D_\Omega$, 
we obtain the following.
\begin{equation}
f(1)
\le
\frac{C_\eta}{\beta^{3/2+\eta/2}}
\frac{A(\partial\Omega)}{V(\Omega)} D_\Omega^\eta
\int_{0}^{D_\Omega} e^{-z^{2}/(C\,\beta)}\,\dd z.
\end{equation}
Moreover, we have,
\begin{equation}
D_\Omega^{\eta}
\int_{0}^{D_\Omega} e^{-z^{2}/(C\,\beta)}\,\dd z
<
D_\Omega^{\eta}
\int_{0}^{\infty} e^{-z^{2}/(C\,\beta)}\,\dd z
=
D_\Omega^{\eta}\sqrt{\frac{\pi\,C\,\beta}{2}}.
\end{equation}
This yields
\begin{equation}
f(1)
\le
\frac{C_\eta}{\beta^{1+\eta/2}}
\frac{A(\partial\Omega)}{V(\Omega)}
D_\Omega^{\eta}.
\end{equation}
Therefore, the Euler--Maclaurin contribution
\(
\frac{f(\infty)-f(1)}{2}
\)
is again bounded at the same order.

\paragraph{Third Euler--Maclaurin term.}

We now estimate the third Euler--Maclaurin contribution
\begin{equation}
\mathcal{R}
:=
\int_{1}^{\infty} f'(k)\left(k-\lfloor k\rfloor-\frac12\right)\dd k
=
\int_{1}^{\infty} f'(k)\,\psi(k)\,\dd k,
\end{equation}
where
\begin{equation}
\psi(k):=k-\lfloor k\rfloor-\frac12,\qquad|\psi(k)|\le \frac12.
\end{equation}
Introduce
\begin{equation}
g(k)
:=
\beta^{-\eta/4}\frac{e^{-z^{2}/(C\,k\beta)}\,e^{-k\beta a}}{(k\beta)^{\alpha}},
\quad
\alpha:=\frac{3}{2}+\frac{\eta}{4}\quad
\text{so that}\quad
f(k)=\int_{\Omega}\frac{\dd X}{V(\Omega)}\;z^{\eta}\,g(k).
\end{equation}
Then
\begin{equation}
f'(k)=\int_{\Omega}\frac{\dd X}{V(\Omega)}\;z^{\eta}\,g'(k).
\end{equation}
%We compute the logarithmic derivative:
%\begin{equation}
%\ln g(k)
%=
%-\frac{z^{2}}{k\beta}-k\beta a-\alpha\ln(k\beta),
%\end{equation}
%hence
%\begin{equation}
%\big(\ln g(k)\big)'
%=
%\frac{z^{2}}{k^{2}\beta}-\beta a-\frac{\alpha}{k}
%=
%\frac{g'(k)}{g(k)}.
%\end{equation}
%Therefore,
%\begin{equation}
%g'(k)
%=
%\left(
%\frac{z^{2}}{k^{2}\beta}-\beta a-\frac{\alpha}{k}
%\right)
%\frac{e^{-z^{2}/(k\beta)}\,e^{-k\beta a}}{(k\beta)^{\alpha}},
%\end{equation}
%and
%\begin{equation}
%f'(k)
%=
%\int_{\Omega}\frac{\dd X}{V(\Omega)}\;z^{\eta}
%\left(
%\frac{z^{2}}{k^{2}\beta}-\beta a-\frac{\alpha}{k}
%\right)
%\frac{e^{-z^{2}/(k\beta)}\,e^{-k\beta a}}{(k\beta)^{\alpha}}.
%\end{equation}
We estimate the third Euler--Maclaurin term as
\begin{equation}
|\mathcal{R}|
\le
C_\eta\,\beta^{-\eta/4}\int_{1}^{\infty}\!\dd k
\int_{\Omega}\frac{\dd X}{V(\Omega)}\;z^{\eta}
\left|
-\beta a-\frac{\alpha}{k}+\frac{z^{2}}{C\,k^{2}\beta}
\right|
\frac{e^{-z^{2}/(C\,k\beta)}\,e^{-k\beta a}}{(k\beta)^{\alpha}}|\psi(k)|.
\end{equation}
\textbf{1st term (the $-\beta a$ contribution):}
Keeping only the contribution proportional to $\beta a$ we bound
\begin{equation}
\mathcal{R}_1
\le
C_\eta\, \beta^{-\eta/4}\int_{1}^{\infty}\!\dd k
\int_{\Omega}\frac{\dd X}{V(\Omega)}\;z^{\eta}\;
\frac{a}{\beta^{\alpha-1}k^{\alpha}}\;
e^{-z^{2}/(C\,k\beta)}\,e^{-k\beta a}\;|\psi(k)|.
\end{equation}
Write $k=n+\zeta$ with $n\in\mathbb{N}$ and $\zeta\in[0,1]$, so that $\dd k=\dd\zeta$ and
$\psi(k)=\zeta-\tfrac12$. Using the estimates
\begin{equation}
|\zeta-\tfrac12|\le \tfrac12,\quad
e^{-z^{2}/(C\,\beta(n+\zeta))}\le e^{-z^{2}/(C\,\beta(n+1))},\quad
e^{-\beta a(n+\zeta)}\le e^{-\beta an},\quad
(n+\zeta)^{-\alpha}\le n^{-\alpha}
\end{equation}
we obtain
\begin{equation}
\mathcal{R}_1
\le
C_\eta\,\beta^{-\alpha+1-\eta/4}a
\int_{\Omega}\frac{\dd X}{V(\Omega)}\;z^{\eta}
\sum_{n=1}^{\infty}\int_{0}^{1}\dd\zeta\;
\frac{e^{-z^{2}/(C\,\beta(n+1))}\,e^{-\beta an}}{2\,n^{\alpha}}.
\end{equation}
Since $\frac{1}{n+1}\le \frac12$ for $n\ge 1$, we have
$e^{-z^{2}/(C\,\beta(n+1))}\le e^{-z^{2}/(2C\,\beta)}$, hence
\begin{equation}
\mathcal{R}_1
\le
C_\eta\,\beta^{-\alpha+1-\eta/4}a
\int_{\Omega}\frac{\dd X}{V(\Omega)}\;z^{\eta}\,e^{-z^{2}/(2C\,\beta)}
\sum_{n=1}^{\infty}\frac{e^{-\beta an}}{n^{\alpha}}.
\end{equation}
The sum is convergent. Before passing to the boundary normal coordinates, noting that we estimate the variable $z$ by its possible maximum $D_\Omega$ and then   similarly estimating
$z^{\eta}\le D_\Omega^{\eta}$ on $[0,D_\Omega]$, we get (employing convex volume estimate)
\begin{equation}
\mathcal{R}_1
\le
C_\eta\,\beta^{-\eta/2}a\,
\frac{A(\partial\Omega)}{V(\Omega)}\,
D_\Omega^{\eta}.
\end{equation}
{\bf Remark:}  $\beta$ behavior of our bound  is actually better, one can see that the sum is dominated by $e^{-\beta a} C$ as $\sum {1\over n^{3/2}}$ is convergent.\\
\textbf{2nd term (the $-\alpha/k$ contribution):}
For the term proportional to $\alpha/k$ we similarly bound
\begin{equation}
\mathcal{R}_2
\le
C_\eta\,\beta^{-\eta/4}\int_{1}^{\infty}\!\dd k
\int_{\Omega}\frac{\dd X}{V(\Omega)}\;z^{\eta}\;
\frac{\alpha}{\beta^{\alpha}k^{\alpha+1}}\;
e^{-z^{2}/(C\,k\beta)}\,e^{-k\beta a}\;|\psi(k)|.
\end{equation}
With the same change $k=n+\zeta$ and the same estimates as above, we obtain
\begin{equation}
\mathcal{R}_2
\le
C_\eta\,\beta^{-\alpha-\eta/4}
\int_{\Omega}\frac{\dd X}{V(\Omega)}\;z^{\eta}\,e^{-z^{2}/(2C\,\beta)}
\sum_{n=1}^{\infty}\frac{e^{-\beta an}}{n^{\alpha+1}}.
\end{equation}
The sum is convergent, and passing to boundary normal coordinates (noting it cannot be  extended beyond $D_\Omega$) gives
\begin{equation}
\mathcal{R}_2
\le
C_\eta\,\beta^{-1-\eta/2}\,
\frac{A(\partial\Omega)}{V(\Omega)}\,
D_\Omega^{\eta}.
\end{equation}
% The remaining (third) contribution proportional to z^2/(k^2 beta)
% is set up above and can be bounded similarly once needed.
\textbf{3rd term (the $z^2/(C\,k^2\beta)$ contribution):} Keeping only the contribution proportional to $\frac{z^{2}}{C\,k^{2}\beta}$, we obtain
\begin{equation}
|\mathcal{R}_3|
\le
C_\eta\,\beta^{-\eta/4}
\int_{1}^{\infty}\dd k
\int_{\Omega}\frac{\dd X}{V(\Omega)}\;
z^{\eta+2}
\frac{e^{-z^{2}/(C\,k\beta)}\,e^{-k\beta a}}{(k)^{\alpha+2}(\beta)^{\alpha+1}}\, |\psi(k)| .
\end{equation}
Writing $k=n+\zeta$ with $n\in\mathbb{N}$ and $\zeta\in[0,1]$, and using
\[
e^{-z^{2}/(C\,\beta(n+\zeta))}\le e^{-z^{2}/(C\,\beta(n+1))},
\quad
e^{-\beta a(n+\zeta)}\le e^{-\beta an},
\quad
(n+\zeta)^{-(\alpha+2)}\le n^{-(\alpha+2)},
\]
where $\alpha=3/2+\eta/4$
we find
\begin{equation}
|\mathcal{R}_3|
\le
C_\eta\,\beta^{-\eta/4}
\int_{\Omega}\frac{\dd X}{V(\Omega)}\;
\sum_{n=1}^{\infty}
\frac{e^{-n\beta a}}{n^{\alpha+2}}\;
z^{\eta+2}\,
e^{-z^{2}/(C\,\beta(n+1))}.
\end{equation}
We now split the exponential as
\[
e^{-z^{2}/(C\,\beta(n+1))}
=
e^{-z^{2}/(2C\,\beta(n+1))}\,
e^{-z^{2}/(2C\,\beta(n+1))}
\]
and define
\[
h(z):=z^{2}e^{-z^{2}/(2C\,\beta(n+1))},
\]
The function $h(z)$ attains its maximum at
\(
z^{2}=2C\,\beta(n+1)
\),
with
\(
h(z)\le 2C\,\beta(n+1)e^{-1}
\).
Therefore,
\begin{equation}
z^{\eta+2}e^{-z^{2}/(C\beta(n+1))}
\le
C_\eta\,\beta(n+1)\,z^{\eta}e^{-z^{2}/(2C\beta(n+1))}.
\end{equation}
Substituting this bound and using the estimate
\[
e^{-z^{2}/(2C\,\beta(n+1))}\le e^{-z^{2}/(4nC\,\beta)},
\]
which follows from \(n+1\ge 2\), we obtain
%\begin{equation}
%|\mathcal{R}_3|
%\le
%C_\eta
%\int_{\Omega}\frac{\dd X}{V(\Omega)}\;
%\sum_{n=1}^{\infty}
%\frac{(n+1)e^{-n\beta a}}{n^{\alpha+2}}\;
%z^{\eta}\,
%e^{-z^{2}/(2\beta(n+1))}.
%\end{equation}
\begin{equation}
|\mathcal{R}_3|
\le
C_\eta\,\beta^{-\eta/4}\sum_{n=1}^{\infty}
\int_{\Omega}\frac{\dd X}{V(\Omega)}\;
z^{\eta}e^{-z^{2}/(4nC\,\beta)}
\frac{(n+1)e^{-n\beta a}}{n^{\alpha+2}}.
\end{equation}
Before passing to boundary normal coordinates, we use
$z^{\eta}\le D_\Omega^{\eta}$ for $0\le z\le D_\Omega$, we obtain
\begin{equation}
\int_{\Omega}\dd X\;z^{\eta}e^{-z^{2}/(4nC\,\beta)}
\le
A(\partial\Omega)\,D_\Omega^{\eta}
\int_{0}^{\infty}e^{-z^{2}/(4nC\,\beta)}\,\dd z
=
A(\partial\Omega)\,D_\Omega^{\eta}\,\sqrt{\pi nC\,\beta}.
\end{equation}
Since $\sum_{n\ge1} n^{-(\alpha+1/2)} e^{-\beta a n}$ is convergent, we obtain 
\begin{equation}
|\mathcal{R}_3|
\le
C_\eta\,
\frac{A(\partial\Omega)}{V(\Omega)}\,
D_\Omega^{\eta}\,
\beta^{-(1+\eta/2)}.
\end{equation}
Note that similar remarks apply to the summations, they are bounded by $e^{-\beta a}$-terms so they have a better convergence regarding the temperature limits.

\subsubsection*{Second Term of the Finite--Temperature Depletion Coefficient}
We start from
\begin{align}
\tilde n^{(2)}_e(T)
&=
\sum_{k=1}^{\infty}
a \int_0^\infty dt\;
\frac{1}{V(\Omega)}
\Tr\!\left(e^{-\Delta\sqrt{(k\beta)^2+t^2}}\right)
e^{-a\sqrt{(k\beta)^2+t^2}}
I_1(at).
\end{align}
Substituting the bound yields
\begin{align}
\tilde n^{(2)}_e(T)
\le
a C_\eta
\int_\Omega \frac{\dd X}{V(\Omega)}\;
z^\eta
\sum_{k=1}^{\infty}
\int_0^\infty \dd t\;
\frac{
e^{-z^2/C\,\sqrt{(k\beta)^2+t^2}}
}{
\big(\sqrt{(k\beta)^2+t^2}\big)^{3/2+\eta/2}
}
e^{-a\sqrt{(k\beta)^2+t^2}}
I_1(at).
\end{align}
Since the integrand is non–negative and absolutely convergent, we may exchange the sum and the
$t$–integration:
\begin{align}
\tilde n^{(2)}_e(T)
\le
a C_\eta
\int_\Omega \frac{\dd X}{V(\Omega)}\;
z^\eta
\int_0^\infty \dd t\; I_1(at)
\sum_{k=1}^{\infty} f(k),
\end{align}
where we define
\begin{equation}\label{fofk}
f(k)
:=
\frac{
\exp\!\left(-\frac{z^2}{C\,\sqrt{(k\beta)^2+t^2}}\right)
}{
\big[(k\beta)^2+t^2\big]^{3/4+\eta/4}
}
\,
e^{-a\sqrt{(k\beta)^2+t^2}}.
\end{equation}

We apply the Euler--Maclaurin formula
\begin{equation}
\sum_{k=1}^{\infty} f(k)
=
\int_{1}^{\infty} f(k)\,\dd k
+
\frac{f(\infty)-f(1)}{2}
+
\int_{1}^{\infty} f'(k)\left(k-\lfloor k\rfloor-\frac12\right)\dd k.
\end{equation}
Accordingly, we decompose
\begin{equation}
\tilde n^{(2)}_e(T)
\le
\tilde n^{(2)}_{e,1}(T)
+
\tilde n^{(2)}_{e,2}(T)
+
\tilde n^{(2)}_{e,3}(T),
\end{equation}
where
\begin{align}
\tilde n^{(2)}_{e,1}(T)
&:=
a C_\eta
\int_{\Omega}\frac{\dd X}{V(\Omega)}\, z^\eta
\int_{0}^{\infty}\dd t\; I_1(at)
\int_{1}^{\infty} f(k)\,\dd k,
\\[4pt]
\tilde n^{(2)}_{e,2}(T)
&:=
a C_\eta
\int_{\Omega}\frac{\dd X}{V(\Omega)}\, z^\eta
\int_{0}^{\infty}\dd t\; I_1(at)\;
\frac{f(\infty)-f(1)}{2},
\\[4pt]
\tilde n^{(2)}_{e,3}(T)
&:=
a C_\eta
\int_{\Omega}\frac{\dd X}{V(\Omega)}\, z^\eta
\int_{0}^{\infty}\dd t\; I_1(at)
\int_{1}^{\infty} f'(k)\left(k-\lfloor k\rfloor-\frac12\right)\dd k.
\end{align}

\paragraph{First Euler--Maclaurin term.}

We start from the first Euler--Maclaurin contribution, Since the integrand is non--negative, we may extend $1\to 0$:
\begin{align}
\tilde n^{(2)}_{e,1}(T)
\le
a C_\eta
\int_{\Omega}\frac{\dd X}{V(\Omega)}\, z^\eta
\int_{0}^{\infty}\dd t\; I_1(at)
\int_{0}^{\infty}\dd k\;
\frac{
\exp\!\left(-\frac{z^2}{C\,\sqrt{(k\beta)^2+t^2}}\right)
}{
\big[(k\beta)^2+t^2\big]^{3/4+\eta/4}
}
\,
e^{-a\sqrt{(k\beta)^2+t^2}}.
\end{align}
Setting $t=\rho\cos\theta$ and $k\beta=\rho\sin\theta,$ gives;
\begin{align}
\tilde n^{(2)}_{e,1}(T)
\le
\frac{a C_\eta}{\beta}
\int_{\Omega}\frac{\dd X}{V(\Omega)}\, z^\eta
\int_{0}^{\infty}\dd\rho\;
\frac{e^{-z^{2}/(C\,\rho)}\,\rho\, e^{-a\rho}}{\rho^{\frac32+\frac{\eta}{2}}}
\int_{0}^{\pi/2}\dd\theta\; I_1\!\big(a\rho\cos\theta\big).
\end{align}
Taking the $\theta$ integral with the formula in Appendix [\ref{preliminaryI}] gives;
\begin{align}\label{ne1}
\tilde n^{(2)}_{e,1}(T)
\le
\frac{ C_\eta}{\beta}
\int_{\Omega}\frac{\dd X}{V(\Omega)}\, z^\eta
\int_{0}^{\infty}\dd\rho\;
\frac{e^{-z^{2}/(C\,\rho)}\,e^{-a\rho}}{\rho^{\frac32+\frac{\eta}{2}}}\,
\sinh^{2}\!\Big(\frac{a\rho}{2}\Big).
\end{align}
We split the $\rho$--integral into two pieces:
\begin{align}
\int_{0}^{\infty}\dd\rho\;
\frac{e^{-z^{2}/(C\,\rho)}\,e^{-a\rho}}{\rho^{\frac32+\frac{\eta}{2}}}\,
\sinh^{2}\!\Big(\frac{a\rho}{2}\Big)
=
\int_{0}^{2/a}(\cdots)\,\dd\rho
+
\int_{2/a}^{\infty}(\cdots)\,\dd\rho
=: I_{\mathrm{small}}+I_{\mathrm{large}}.
\end{align}
Using the elementary bound for $\sinh$; For $0<x<1$ i.e. $\sinh(x) \le x+\frac{x^{3}}{5}+\ldots$
we get, $\sinh^{2}\!\Big(\frac{a\rho}{2}\Big)
\le \frac{9}{25}a^2\rho^{2}$. Then, one has;
\begin{align}
I_{\mathrm{small}}
&\le
\frac{9}{25}a^{2}\int_{0}^{2/a}\dd\rho\;
\frac{e^{-z^{2}/(C\,\rho)}\,e^{-a\rho}}{\rho^{\frac32+\frac{\eta}{2}}}\,
\rho^{2}
=
\frac{9}{25}a^{2}\int_{0}^{2/a}\dd\rho\;
\rho^{\frac12-\frac{\eta}{2}}\,e^{-z^{2}/(C\,\rho)}\,e^{-a\rho}.
\label{eq:Ismall-TODO-start}
\end{align}
For $0<\rho\le 2/a$ we have $\rho^{-1}\ge a/2$, hence $e^{-z^{2}/C\,\rho}\le e^{-(az^{2}/2C)}.$
Therefore, from \eqref{eq:Ismall-TODO-start},
\begin{align}
I_{\mathrm{small}}
&\le
\frac{9}{25}a^{2}\,e^{-(az^{2}/2C)}
\int_{0}^{2/a}\rho^{\frac12-\frac{\eta}{2}}\,e^{-a\rho}\,\dd\rho.
\end{align}
We make the change of variables $u=a\rho$, to obtain
\begin{align}
\int_{0}^{2/a}\rho^{\frac12-\frac{\eta}{2}}\,e^{-a\rho}\,\dd\rho
&=
a^{-\left(\frac32-\frac{\eta}{2}\right)}
\int_{0}^{2}u^{\frac12-\frac{\eta}{2}}e^{-u}\,\dd u
\le
a^{-\left(\frac32-\frac{\eta}{2}\right)}
\int_{0}^{\infty}u^{\frac12-\frac{\eta}{2}}e^{-u}\,\dd u \nonumber\\
&=
a^{-\left(\frac32-\frac{\eta}{2}\right)}
\Gamma\!\Big(\frac32-\frac{\eta}{2}\Big),
\end{align}
which is finite for $0<\eta<1$. Using $z^\eta\le D_\Omega^\eta$ for $0\le z\le D_\Omega$ and passing to
boundary normal coordinates,
\begin{align}
\int_{\Omega}\dd X\; z^\eta e^{-(az^{2}/2C)}
&\le
A(\partial\Omega)\, D_\Omega^\eta\int_{0}^{D_\Omega}  e^{-(az^{2}/2C)}\,\dd z
\le
A(\partial\Omega)\,D_\Omega^\eta
\int_{0}^{\infty} e^{-(az^{2}/2C)}\,\dd z \nonumber\\
&=
A(\partial\Omega)\,D_\Omega^\eta\sqrt{\frac{\pi\,C}{2a}}.
\end{align}
%Substituting into \eqref{ne1}, we conclude
%\begin{equation}
%\tilde n^{(2)}_{e,1}(T)
%\le
%a^{\eta/2}\frac{C_\eta}{\beta}\,
%\frac{A(\partial\Omega)}{V(\Omega)}\,
%D_\Omega^\eta\,.
%\label{eq:ne21-final}
%\end{equation}
On the other hand, the large--$\rho$ contribution ($\rho\ge 2/a$) can be estimated using the dominant term of $\sinh^2(a\rho/2)$, namely $e^{a\rho}$.
\begin{eqnarray}
I_{\mathrm{large}}
&&\le
\int_{2/a}^{\infty}\dd\rho\;
\frac{e^{-z^{2}/C\,\rho}}{\rho^{\frac32+\frac{\eta}{2}}} \nonumber \\
&&\leq (a/2)^{1/2+\eta/2} \int_1^\infty \dd \xi {e^{-{az^2\over 2C\xi}}\over \xi^{3/2}}\nonumber \\
&&\leq{(a/2)^{1/2+\eta/2}\over 4}\int_0^\infty \dd \xi\, {e^{-{az^2\over 2C\xi}}\over \xi^{3/2}}={(a/2)^{\eta/2}\over 4}{\sqrt{C}\over z}.
\end{eqnarray}
Substituting both estimates into \eqref{ne1},
\begin{equation}
    C_{\eta} a^{\eta/2}\int {\dd X\over V(\Omega)} {z^\eta\over z}\leq C_{\eta} a^{\eta/2} {A(\partial \Omega)\over V(\Omega)} D_\Omega^\eta. 
\end{equation}

\paragraph{Second Euler--Maclaurin term.}

The second Euler--Maclaurin contribution is given by
\begin{equation}
\tilde n^{(2)}_{e,2}(T)
:=
a C_\eta
\int_{\Omega}\frac{\dd X}{V(\Omega)}\, z^\eta
\int_{0}^{\infty}\dd t\; I_1(at)\;
\frac{f(\infty)-f(1)}{2},
\end{equation}
where $f(k)$ is as given in \ref{fofk}. Note that $f(k)\to 0$ as $k\to\infty$, hence $f(\infty)=0.$
Moreover,
\begin{equation}
f(1)
=
\frac{
e^{-{z^{2}}/C{\sqrt{\beta^{2}+t^{2}}}}
}{
(\beta^{2}+t^{2})^{\frac34+\frac{\eta}{4}}
}
e^{-a\sqrt{\beta^{2}+t^{2}}}
\ge 0.
\end{equation}
Therefore, only the term $-f(1)/2$ remains, and we need to estimate its absolute value. For this purpose, observe that 
$I_1(x)<I_{1/2}(x)$, a well-known inequality for $x>0$ \cite{Cochran1967}.
Note that $I_{1/2}(x)=\sqrt{2\over \pi x} \sinh(x)$, hence absolute value of the expression above is less than
the one for which  $I_1$ replaced with $I_{1/2}$. 
Scaling the variable $t$ with $\beta$ and using the naive inequality $\sinh(a\beta t)e^{-a\beta(1+t)^{1/2}}<1$, we obtain
\begin{eqnarray}
   &&{C_\eta\over \beta^{2+\eta/2}} \int {\dd X\over V(\Omega)} z^\eta \int_0^\infty  \dd t {e^{-{z^2/C\, \beta(1+t^2)^{1/2}}} \over [1+t^2]^{3/4+\eta/4}}\nonumber \\
   &&\leq {C_\eta D^\eta_\Omega\over \beta^{2+\eta/2}} \int_0^\infty  \Big[\int_0^\infty \dd z\,e^{-{z^2/C\,\beta(1+t^2)^{1/2}}}\Big] {\dd t\over t^{1/2} [1+t^2]^{3/4+\eta/4}}
,\end{eqnarray}
where we first interchange the $t$– and $z$–integrations and then extend the upper
limit of the $z$–integration to $\infty$.
The resulting $z$–integration produces the factor
$
\beta^{-3/2-\eta/2}
\int_{0}^{\infty}
\frac{\dd t}{t^{1/2}\,[1+t^{2}]^{1/2+\eta/4}},
$
which is convergent thanks to the additional $\eta/4$ in the exponent.
This yields a bound of the desired form.
%%%%%%%%%%%%%%%%%%%%%%%%%%%%%%%%%%%%%%%%%%%%%%%%%%%%%%%%%%%%
% 3rd Euler--Maclaurin contribution (Second depletion term)
% (only the first two handwritten pages)
%%%%%%%%%%%%%%%%%%%%%%%%%%%%%%%%%%%%%%%%%%%%%%%%%%%%%%%%%%%%
\paragraph{Third Euler--Maclaurin term.}

\begin{equation}
\tilde n^{(2)}_{e,3}(T)
=
a\,C_\eta
\int_{\Omega}\frac{\dd X}{V(\Omega)}\,z^{\eta}
\int_{0}^{\infty}\dd t\; I_{1}(a t)
\int_{1}^{\infty}\dd k\;
\frac{\dd}{\dd k}\!\big[f(k,t)\big]\,
\Big(k-\lfloor k\rfloor-\frac12\Big).
\end{equation}
We immediately notice that $f(k,t)$ is symmetric in $k,t$, we thus exploit this employing the following sequence of changes:
\begin{equation}
\frac{\dd}{\dd k}
\;\longrightarrow\;
\beta\,\frac{\dd}{\dd(k\beta)}
\;\longrightarrow\;
\beta\,\frac{\dd}{\dd t}.
\end{equation}
As a result, we find that 
\begin{align}
\tilde n^{(2)}_{e,3}(T)
&=
a\,C_\eta
\int_{\Omega}\frac{\dd X}{V(\Omega)}\,z^{\eta}
\int_{1}^{\infty}\dd k\;\psi(k)
\int_{0}^{\infty}\dd t\;
\Bigg[
\frac{\dd}{\dd t}\Big(I_{1}(a t)\,f(k,t)\Big)
-
f(k,t)\,\frac{\dd}{\dd t}I_{1}(a t)
\Bigg],
\end{align}
As usual, we introduce $\psi(k)$ to represent the saw-tooth function, we note that $|\psi(k)|\leq\frac12 $ to get a simpler estimate;
\begin{align}
\tilde n^{(2)}_{e,3}(T)
&<
a\,C_\eta
\int_{\Omega}\frac{\dd X}{V(\Omega)}\,z^{\eta}
\int_{1}^{\infty}\dd k\;\int_{0}^{\infty}\dd t\;
f(k,t)\,I_1'(a t).
\end{align}
\noindent\textbf{Note that} we employ in the integration by parts argument the fact that,
\begin{align}
\Big[I_1(a t)\,f(k,t)\Big]_{0}^{\infty}
&=
\left[
I_1(a t)\cdot
\frac{
e^{-z^{2}/C\,\sqrt{(k\beta)^{2}+t^{2}}}\times
e^{-a\sqrt{(k\beta)^{2}+t^{2}}}
}{
\big[(k\beta)^{2}+t^{2}\big]^{\frac34+\frac{\eta}{4}}
}
\right]_{0}^{\infty}
=0,
\end{align}
which means there is no boundary contribution.

\noindent We now use the identity,
\begin{equation}
\frac{\dd}{\dd x} I_{\nu}(x)
=
\frac12\Big(I_{\nu-1}(x)+I_{\nu+1}(x)\Big) \quad \text{to get}\quad\frac{\dd}{\dd t}I_1(a t)
=
\frac{a}{2}\Big(I_{0}(a t)+I_{2}(a t)\Big).
\end{equation}
Now note that $I_0>I_2$, thus
\begin{align}
\tilde n^{(2)}_{e,3}(T)
&<
a^{2}C_\eta
\int_{\Omega}\frac{\dd X}{V(\Omega)}\,z^{\eta}
\int_{0}^{\infty}\dd k
\int_{0}^{\infty}\dd t\;
\frac{
e^{-z^{2}/C\,\sqrt{(k\beta)^{2}+t^{2}}}\times
e^{-a\sqrt{(k\beta)^{2}+t^{2}}}
}{
\big[(k\beta)^{2}+t^{2}\big]^{\frac34+\frac{\eta}{4}}
}\;
I_{0}(a t).
\end{align}
Using the same coordinate transformation as in the previous case, namely $t=\rho\cos\theta$ and $k\beta=\rho\sin\theta$, we obtain
\begin{align}
\tilde n^{(2)}_{e,3}(T)
&\le
\frac{a^{2}C_\eta}{\beta V(\Omega)}
\int_{\Omega}\dd X\,z^{\eta}
\int_{0}^{\infty}\dd\rho
\int_{0}^{\pi/2}\dd\theta\;
\frac{\rho}{\rho^{\frac32+\frac{\eta}{2}}}\;
e^{-z^{2}/(C\,\rho)}\,
e^{-a\rho}\,
I_{0}\!\big(a\rho\cos\theta\big).
\end{align}
We use the product formula for modified Bessel functions collected in
Appendix~\ref{app:I0_product_formula} to take the $\dd \theta$ integral.
\begin{align}
\tilde n^{(2)}_{e,3}(T)
&\le
\frac{a^{2}C_\eta}{\beta V(\Omega)}
\int_{\Omega}\dd X\,z^{\eta}
\int_{0}^{\infty}\dd\rho\;\rho\,
\frac{
e^{-z^{2}/(C\,\rho)}\times e^{-a\rho}
}{
\rho^{\frac32+\frac{\eta}{2}}
}\;
\Big[I_{0}\!\Big(\frac{a\rho}{2}\Big)\Big]^{2}.
\end{align}
Split $\big[I_0\big]^2$ behaviour into two pieces:
\begin{align}
\int_{0}^{\infty} \dd \rho\;
\rho \,
\frac{e^{{\frac{-z^2}{C\,\rho}} - a\rho}}{\rho^{\frac32+\frac{\eta}{2}}}
\Big[I_0(a\rho/2)\Big]^2
&=
\int_{0}^{2/a}
\frac{e^{{\frac{-z^2}{C\,\rho}} - a\rho}}{\rho^{\frac12+\frac{\eta}{2}}}
\cdot \Big[I_0(\text{small})\Big]^2
\; \dd\rho
\nonumber\\
&\quad
+
\int_{2/a}^{\infty}
\frac{e^{{\frac{-z^2}{C\,\rho}} - a\rho}}{\rho^{\frac12+\frac{\eta}{2}}}
\Big[I_0(\text{large})\Big]^2
\; \dd\rho .
\end{align}
$I_0(\text{small})$ and $I_0(\text{large})$ behaviours are discussed in the appendix \ref{app:I0_global}. After inserting the relevant estimations we get;
\begin{eqnarray}
I_{\mathrm{small}}
\;\le\;
\frac{a^{2} C_\eta}{\beta V(\Omega)}
\int_{\Omega} \dd X\, z^{\eta}\, e^{-z^{2} a/2C}
\;\le\;
\frac{a^{2} C_\eta A(\partial\Omega)}{\beta V(\Omega)}
\int_{0}^{D_{\Omega}} \dd z\, z^{\eta}\, e^{-z^{2} a/2C}\nonumber \\
\;\le\;
\frac{a^{2} C_\eta A(\partial\Omega)\,D_{\Omega}^{\eta}}{\beta V(\Omega)}
\int_{0}^{\infty} \dd z\, e^{-z^{2} a/2C}\le\;\frac{a^{3/2} C_\eta}{\beta}\frac{ A(\partial\Omega)\,}{V(\Omega)}D_{\Omega}^{\eta}
\end{eqnarray}
For the large integral piece;
\begin{align}
&\int_{2/a}^{\infty}\dd\rho\;
\frac{e^{{\frac{-z^2}{C\,\rho}} - a\rho}}{\rho^{\frac12+\frac{\eta}{2}}}\,
\frac{e^{a\rho}}{a\rho}
=
\frac{1}{a}\int_{2/a}^{\infty}\dd\rho\;
\frac{e^{-z^{2}/C\rho}}{\rho^{\frac32+\frac{\eta}{2}}}\le
\left(\frac{a}{2}\right)^{\frac{\eta}{2}}\frac{1}{a}
\int_{2/a}^{\infty}\dd\rho\;
\frac{e^{-z^{2}/C\rho}}{\rho^{\frac32}}
\nonumber\\
&\le
\frac{2}{a}\left(\frac{a}{2}\right)^{\frac{\eta+3}{2}}
\int_{0}^{\infty}\dd\xi\;
\frac{e^{-a z^{2}/2C\xi}}{\xi^{\frac32}}=
\frac{\sqrt{2C\pi}}{z\,a^{1/2}}
\left(\frac{a}{2}\right)^{\frac{\eta}{2}} .
\end{align}
Therefore;
\begin{align}
I_{\mathrm{large}}\
&\le
\frac{a^{1+\eta/2}\,C_{\eta}}{\beta}\;
\frac{A(\partial\Omega)}{V(\Omega)}
\int_{0}^{D_{\Omega}}\dd z\; z^{\eta-1}
\nonumber\\
&<
\frac{a^{1+\eta/2}\,C_{\eta}}{\beta}\;
\frac{A(\partial\Omega)}{V(\Omega)}\;
D_{\Omega}^{\eta}.
\end{align}
%%%%%%%%%%%%%%%%%%%%%%%%%%%%%%%%%%%%%%%%%%%%%%%%%%%%%%%%%%%%
\section*{Chemical Potential}
In the Bogoliubov model we note that $a=u_0 n_0$ with $n_0=N/V(\Omega)$. The chemical potential in the condensed phase is defined as
\begin{equation}\label{mu_def}
\mu:=\frac{\partial E_{gr}}{\partial N}.
\end{equation}
We differentiate at fixed domain $\Omega$, fixed coupling $u_0$ and temperature.
Since $a=u_0 n_0$, differentiation with respect to $N$ reduces to differentiation
with respect to $a$,
\begin{equation}\label{dNdA}
\frac{\partial}{\partial N}
=\frac{u_0}{V(\Omega)}\,\frac{\partial}{\partial a}.
\end{equation}
We introduce;
\begin{equation}\label{I_def}
\mathcal{I}(a)
:=\int_{0}^{\infty} \dd \tau\int_{0}^{1} \dd x\;
\sqrt{1-x^{2}}\,
\Big[(1-e^{-a\tau(1-x)})+(1-e^{-a\tau(1+x)})\Big]\;
{\rm Tr}'\,e^{\tau \Delta}.
\end{equation}
Its derivative is
\begin{equation}\label{Iprime}
\mathcal{I}'(a)
=
\int_{0}^{\infty} \dd t\int_{0}^{1} \dd x\;
t\,\sqrt{1-x^{2}}\,
\Big[(1-x)e^{-a t(1-x)}+(1+x)e^{-a t(1+x)}\Big]\;
{\rm Tr}'\,e^{t\Delta}.
\end{equation}
Using the ground--state energy formula given above (\ref{GSnrg}), we may write
\begin{equation}\label{mu_split}
\mu
= u_0 n_0
+\frac{u_0}{V(\Omega)}\Big[\frac{a}{\pi}\mathcal{I}(a)
+\frac{a^2}{2\pi}\mathcal{I}'(a)\Big].
\end{equation}
Define
\begin{equation}\label{F_def_mu}
F(t,x):=\sqrt{1-x^2}\Big[(1-e^{-a t(1-x)})+(1-e^{-a t(1+x)})\Big],
\end{equation}
and
\begin{equation}\label{G_def_mu}
G(t,x):=a\sqrt{1-x^2}\Big[(1-x)e^{-a t(1-x)}+(1+x)e^{-a t(1+x)}\Big].
\end{equation}
Then \eqref{mu_split} becomes
\begin{eqnarray}\label{mu_final_formula}
\mu = u_0 n_0 &+&\frac{u_0^2 n_0}{\pi}\, \int_{0}^{\infty} \dd t\int_{0}^{1} \dd x\; F(t,x)\,\frac{1}{V(\Omega)}{\rm Tr}'\,e^{t\Delta}\nonumber\\
&+&\frac{u_0^2 n_0}{2\pi} \int_{0}^{\infty} \dd t\int_{0}^{1} \dd x\; t\,G(t,x)\,\frac{1}{V(\Omega)}{\rm Tr}'\,e^{t\Delta}.
\end{eqnarray}
Let $\mu^{(\infty)}$ denote the infinite--space value obtained by
replacing the Neumann heat kernel trace by the flat space expression.
We estimate
\begin{equation}\label{Delta_mu_def}
|\Delta\mu|
:=|\mu-\mu^{(\infty)}|
\leq |\Delta\mu_{(2)}|+|\Delta\mu_{(3)}|,
\end{equation}
where $|\Delta\mu_{(2)}|$ corresponds to the term with $F$ in \eqref{mu_final_formula}
and $|\Delta\mu_{(3)}|$ corresponds to the term with $t\,G$.
We now apply Brown’s Neumann heat–kernel bound.
\paragraph{Estimate of $|\Delta\mu_{(2)}|$.}
Using equation \ref{replacement}, together with the inequalities,
$e^{-at(1+x)} \le e^{-at(1-x)}$ and $\sqrt{1-x^2}\le 2\sqrt{1-x}$ for $x\in[0,1]$,
which are very similar to those employed in the second section, we derive the following upper bound,
\begin{equation}\label{Dmu2_start}
|\Delta\mu_{(2)}|
\leq
C_{\eta}\,u_0\,a^2
\int_{\Omega}\frac{\dd X}{V(\Omega)}\, z^{\eta}
\int_{0}^{\infty}\dd t\int_{0}^{1} \dd x\int_{0}^{1} \dd \zeta\,(1-x)^{3/2}\,e^{-at(1-x)\zeta}\,\frac{e^{-z^{2}/(C\,t)}}{t^{(1+\eta)/2}}.
\end{equation}
The $t$--integral is of the same type already treated in the previous sections.
Proceeding exactly as before (scaling out $z$ and using the standard Bessel-$K$
representation), one arrives at the bound
\begin{equation}\label{Dmu2_mid}
|\Delta\mu_{(2)}|
\leq
C_{\eta}\,u_0\, a^{1+\eta/4}
\int_{\Omega}\frac{\dd X}{V(\Omega)}\, z^{-1+\eta/2}
\leq
C_{\eta}\,u_0\, a^{1+\eta/4}\frac{A(\partial \Omega)}{V(\Omega)}D_{\Omega}^{\eta/2}
\end{equation}

\paragraph{Estimate of $|\Delta\mu_{(3)}|$.}
We now treat the third term in \eqref{mu_final_formula}. First note that
$e^{-at(1+x)}\leq e^{-at(1-x)}$ for $x\in[0,1]$ and $1-x\leq 1+x\leq 2$,
hence
\begin{equation}\label{G_simple_bound}
0\leq G(t,x)
\leq 4a\sqrt{1-x}\,e^{-at(1-x)}.
\end{equation}
Using \eqref{G_simple_bound} and \eqref{brown} gives
\begin{equation}\label{Dmu3_start}
|\Delta\mu_{(3)}|
\leq
C_{\eta}\,u_0\,a^2
\int_{\Omega}\frac{\dd X}{V(\Omega)}\, z^{\eta}
\int_{0}^{1} \dd x\,\sqrt{1-x}
\int_{0}^{\infty} \dd t\;
\frac{e^{\frac{-z^{2}}{(C\,t)}-a t(1-x)}}{t^{1/2+\eta/2}}.
\end{equation}
The $t$--integral is again of the same model form as in the earlier estimates.
Using the same substitutions as in the previous sections (in particular,
scaling by $z$ and turning the integral into a modified Bessel function and
then using its integral representation to extract the decay), we obtain
\begin{equation}\label{Dmu3_mid}
|\Delta\mu_{(3)}|
\leq
C_{\eta}\,u_0\, a^{1+\eta/4}
\int_{\Omega}\frac{\dd X}{V(\Omega)}\, z^{-1+\eta/2}
\leq
C_{\eta}\,u_0\, a^{1+\eta/4}\,\frac{A(\partial \Omega)}{V(\Omega)}D_{\Omega}^{\eta/2}
\end{equation}
Combining \ref{Dmu2_mid} with \ref{Dmu3_mid} yields,
\begin{equation}
\Delta\mu\leq C_{\eta}\,u_0\, a^{1+\eta/4}\,\frac{A(\partial \Omega)}{V(\Omega)}D_{\Omega}^{\eta/2}.
\end{equation}
The right-hand side vanishes in the thermodynamic limit for a nested family of scaled convex domains $ A(\partial \Omega)/V(\Omega)\sim D_{\Omega}^{-1}$
showing that the chemical potential converges to its bulk value.

\section{Conclusions}

Precise control over the thermodynamic limit has always been an interesting problem. In our present modest contribution, we {\it assume the strict validity of the Bogoliubov} model for a weakly interacting Bose gas and consider scalings of a convex box. We show that for Neumann boundary conditions,  the finite box results, for all the relevant physical quantities,  differ from the infinite volume limit by terms of order $D_\Omega^{-1+\eta}$, where $D_\Omega$ is the diameter and $\eta>0$ is an arbitrarily small constant. This is a strict control over the thermodynamic limit, which is achieved by relatively elementary means. We hope that this result is an initial stage of a deeper investigation. Having completed the present revised version, we became aware of a recent preprint \cite{franklarson} which improves upon Brown's estimates assuming convex regions. Our calculations reveal that, for the estimate we use one cannot take the $\eta\to 0^+$ limit, as this would lead to infinities in the final answer. Physical intuition would suggest that such sensitive dependence of final results on a boundary condition is essentially a technical issue, moreover such a factor has no natural explanation in the theory so it should not be there in the thermodynamic limit. It is an interesting challenge to look for  further improvements  by means of recent developments and perhaps it is possible to  remove the physically ambiguous factor $\eta$ in the final limiting process.

%%%%%%%%%%%%%%%%%%%%%%%%%%%%%%%%%%%%%%%%%%%%%%%%%%%%%%%%%%%%
\appendix

\section{Bounds for the modified Bessel function $I_{0}$}
\label{app:I0_bound}
%%%%%%%%%%%%%%%%%%%%%%%%%%%%%%%%%%%%%%%%%%%%%%%%%%%%%%%%%%%%%%%%%%%%%%%%%%%%%%

In this appendix we collect the bounds on $I_{0}$ that are used in the main
estimates.  The goal is not to optimize constants, but to provide bounds in
two regimes (small and large arguments) in a form suitable for the $\rho$--type
integrals appearing in the finite--temperature depletion terms.

%---------------------------------------------------------------------------
\subsection*{A preliminary identity used in the main text}\label{preliminaryI}
%---------------------------------------------------------------------------

We use the identity
\begin{equation}
\int_{0}^{\pi/2}\dd\theta\; I_1\!\big(a\rho\cos\theta\big)
=
\frac{\pi}{2}\,I_{1/2}\!\Big(\frac{a\rho}{2}\Big)^2,
\end{equation}
together with
\begin{equation}
I_{1/2}(x)
=
\frac{(x/2)^{1/2}}{\sqrt{\pi}\,\Gamma(1)}
\int_{-1}^{1} e^{xt}\,\dd t
=
\sqrt{\frac{2}{\pi x}}\sinh x,
\end{equation}
which yields
\begin{equation}
\int_{0}^{\pi/2}\dd\theta\; I_1\!\big(a\rho\cos\theta\big)
=
\frac{2}{a\rho}\,
\sinh^{2}\!\Big(\frac{a\rho}{2}\Big).
\end{equation}

%---------------------------------------------------------------------------
\subsection*{A product formula used to square $I_0$}
\label{app:I0_product_formula}
%---------------------------------------------------------------------------

We also use the following product formula for modified Bessel functions:
\begin{equation}
I_{\mu}(x)\,I_{\nu}(x)
=
\frac{2}{\pi}\int_{0}^{\pi/2}
I_{\mu+\nu}(2x\cos\theta)\,
\cos\big((\mu-\nu)\theta\big)\,\dd\theta,
\qquad
\text{set }\mu=\nu=0.
\end{equation}
In particular, this yields
\begin{equation}
\big[I_{0}(x)\big]^{2}
=
\frac{2}{\pi}\int_{0}^{\pi/2}
I_{0}(2x\cos\theta)\,\dd\theta,
\qquad
\text{set }x=\frac{a\rho}{2}\ \ \text{for our case.}
\end{equation}

%---------------------------------------------------------------------------
\subsection*{Small--argument control}
\label{app:I0_small}
%---------------------------------------------------------------------------

We recall the series representation
\begin{equation}
I_0(x)
=
\sum_{k=0}^{\infty}
\frac{(x/2)^{2k}}{(k!)^2}
=
1 + \frac{x^2}{2^2} + \frac{x^4}{2^4 (2!)^2} + \cdots
\end{equation}
and note that for $0<x<1$,
\begin{equation}
I_0(x)
<
1 + \frac{1}{2^2} + \frac{1}{2^4 (2!)^2}+...,
\quad
{\rm the\ rapid\ convergence\ implies,}\
\big[I_0(x)\big]^2 < \text{const}.
\label{eq:I0_small_const}
\end{equation}
This crude bound is sufficient for the part of the $\rho$--integrals restricted
to $0\le a\rho \le 2$ (equivalently $0\le \rho \le 2/a$).  In particular, it
justifies treating $\big[I_0(a\rho/2)\big]^2$ as a bounded factor on the
``small--$\rho$'' interval.

\noindent
{\bf Remark.}
For small values of $x$ it is often convenient to keep a quadratic control:
suppose $0\leq x<1$, then a simple estimate reveals that
\[
  I_0(x)\leq 1+ C x^2\, ,
\]
which can be used when one needs a slightly more explicit dependence near the
origin.

%---------------------------------------------------------------------------
\subsection*{Trigonometric bounds}
%---------------------------------------------------------------------------

For $0\le \theta \le \pi/2$, the cosine function satisfies the quadratic bounds
\begin{equation}
1-\frac{\theta^2}{2}
\;\le\;
\cos\theta
\;<\;
1-\frac{\pi^2}{4}\,\theta^2.
\label{eq:cos_bounds}
\end{equation}
Such global bounds are standard and can be found, for instance, in classical
treatments of trigonometric inequalities and in the analysis of oscillatory and
exponential integrals \cite{WatsonBessel,DLMF,CosineBound}.

%---------------------------------------------------------------------------
\subsection*{Integral representation and a global upper bound}
\label{app:I0_global}
%---------------------------------------------------------------------------

We start from the integral representation of $I_0$ (see DLMF~\cite[Eq.~10.32.1]{DLMF}):
\begin{equation}
I_0(x)
=
\frac{1}{\pi}
\int_{0}^{\pi} e^{x\cos\theta}\,\dd\theta,
\qquad x>0.
\label{eq:I0_integral}
\end{equation}
Splitting the integral at $\theta=\pi/2$, we write
\begin{equation}
I_0(x)
=
\frac{1}{\pi}
\left(
\int_{0}^{\pi/2} e^{x\cos\theta}\,\dd\theta
+
\int_{\pi/2}^{\pi} e^{x\cos\theta}\,\dd\theta
\right).
\label{eq:I0_split}
\end{equation}

\paragraph{Estimate of the first contribution.}
For the first term in \eqref{eq:I0_split}, we use the upper bound in
\eqref{eq:cos_bounds}:
\begin{align}
\int_{0}^{\pi/2} e^{x\cos\theta}\,\dd\theta
&\le
\int_{0}^{\pi/2}
\exp\!\left(x-\frac{\pi^2}{4}x\,\theta^2\right)\,\dd\theta
\nonumber\\
&\le
e^{x}
\int_{0}^{\infty}
\exp\!\left(-\frac{\pi^2}{4}x\,\theta^2\right)\,\dd\theta.
\label{eq:first_piece_1}
\end{align}
Evaluating the Gaussian integral,
\begin{equation}
\int_{0}^{\infty} e^{-a\theta^2}\,\dd\theta
=
\frac{1}{2}\sqrt{\frac{\pi}{a}},
\qquad a>0,
\end{equation}
we obtain
\begin{equation}
\int_{0}^{\pi/2} e^{x\cos\theta}\,\dd\theta
\;\le\;
\frac{e^{x}}{\sqrt{\pi x}}.
\label{eq:first_piece}
\end{equation}

\paragraph{Estimate of the second contribution.}
For the second term in \eqref{eq:I0_split}, we set $\theta=\pi-\phi$, which gives
$\cos(\pi-\phi)=-\cos\phi$, and thus
\begin{equation}
\int_{\pi/2}^{\pi} e^{x\cos\theta}\,\dd\theta
=
\int_{0}^{\pi/2} e^{-x\cos\phi}\,\dd\phi.
\end{equation}
Using the lower bound in \eqref{eq:cos_bounds}, we estimate
\begin{align}
\int_{0}^{\pi/2} e^{-x\cos\phi}\,\dd\phi
&\le
\int_{0}^{\pi/2}
\exp\!\left(-x+\frac{x}{2}\phi^2\right)\,\dd\phi
\nonumber\\
&=
e^{-x}
\int_{0}^{\pi/2}
\exp\!\left(\frac{x}{2}\phi^2\right)\,\dd\phi.
\label{eq:second_piece_1}
\end{align}
Introducing the rescaled variable $\phi=\frac{\pi}{2}s$, we find
\begin{equation}
e^{-x}
\int_{0}^{\pi/2}
\exp\!\left(\frac{x}{2}\phi^2\right)\,\dd\phi
=
\frac{\pi}{2}e^{-x}
\int_{0}^{1}
\exp\!\left(\frac{\pi^2}{8}x\,s^2\right)\,\dd s.
\label{eq:second_piece_2}
\end{equation}
Since $0\le s\le 1$ implies $s^2\le s$, this yields
\begin{equation}
\int_{0}^{\pi/2} e^{-x\cos\phi}\,\dd\phi
\;\le\;
\frac{1}{2}
\exp\!\left[x\!\left(\frac{\pi^2}{8}-1\right)\right].
\label{eq:second_piece}
\end{equation}

\paragraph{Resulting global bound.}
Combining \eqref{eq:I0_split}, \eqref{eq:first_piece}, and
\eqref{eq:second_piece}, we arrive at the bound
\begin{equation}
I_0(x)
\;\le\;
\frac{1}{\pi^{3/2}}\frac{e^{x}}{\sqrt{x}}
+
\frac{1}{2}
\exp\!\left[x\!\left(\frac{\pi^2}{8}-1\right)\right].
\label{eq:I0_final_bound}
\end{equation}

%---------------------------------------------------------------------------
\subsection*{Application to $I_0\!\left(\frac{a\rho}{2}\right)$ and a bound for $\big[I_0\big]^2$}
\label{app:I0_square}
%---------------------------------------------------------------------------

Setting $x=\frac{a\rho}{2}$ in \eqref{eq:I0_final_bound}, we obtain
\begin{equation}
I_0\!\left(\frac{a\rho}{2}\right)
\;\le\;
\frac{\sqrt{2}}{\pi^{3/2}}
\frac{e^{a\rho/2}}{\sqrt{a\rho}}
+
\frac{1}{2}
\exp\!\left[\frac{a\rho}{2}\!\left(\frac{\pi^2}{8}-1\right)\right].
\label{eq:I0_arho}
\end{equation}
Squaring this expression and estimating each term separately yields
\begin{equation}
I_0^2\!\left(\frac{a\rho}{2}\right)
\;\le\;
C_1\,\frac{e^{a\rho}}{a\rho}
+
C_2\,\frac{e^{\delta a\rho}}{\sqrt{a\rho}}
+
C_3\,e^{\alpha a\rho},
\label{eq:I0_square}
\end{equation}
where
\begin{equation}
\delta=\frac{\pi^2}{16}\approx0.62,
\qquad
\alpha=\frac{\pi^2}{8}-1\approx0.23,
\end{equation}
and $C_1,C_2,C_3>0$ are numerical constants.

\noindent\textbf{Note that;}
\begin{equation}
\big[I_0(\frac{a\rho}{2})\big]^2
<
C_1\,\frac{e^{a\rho}}{a\rho}
\left[
1
+
\frac{C_2}{C_1}\,
\frac{(a\rho)^{1/2}}{e^{(1-\delta)a\rho}}
+
\frac{C_3}{C_1}\,
\frac{a\rho}{e^{(1-\alpha)a\rho}}
\right].
\end{equation}

\noindent
Let $a\rho=x$.
\begin{equation}
\big[I_0(\frac{x}{2})\big]^2
<
C_1\,\frac{e^{x}}{x}
\left[
1
+
\frac{C_2}{C_1}\,
\frac{\sqrt{x}}{e^{(1-\delta)x}}
+
\frac{C_3}{C_1}\,
\frac{x}{e^{(1-\alpha)x}}
\right].
\end{equation}

\begin{equation}
\frac{2}{a}<\rho<\infty
\quad\Longrightarrow\quad
2<x<\infty .
\end{equation}

\noindent
The last two terms in the bracket are monotone decreasing in $x$ on $[2,\infty)$,
hence their maxima are attained at $x=2$, and therefore
\begin{equation}
\big[I_0(x)\big]^2
\le
C_1\,\frac{e^{x}}{x}
\left[
1 + C_{21} + C_{31}
\right]
\end{equation}
and consequently
\begin{equation}
\big[I_0(x)\big]^2\le C\,\frac{e^{x}}{x}.
\end{equation}

\section{Acknowledgements}

O. T. Turgut thanks M. Asorey and A. Wipf for discussions, and thanks M. Asorey for the invitation to visit the Department of Theoretical Physics of the Zaragoza University where the first version of this  work is completed. O. T. Turgut thanks to G. dell Antonio, L. Dabrowski,  G. Landi and A. Michelangeli for comments on this initial  work and for an opportunity to present his ideas while he was visiting  SISSA.  O. T. Turgut thanks G. Landi for the kind invitation. The authors would like to thank to R. Brown for some clarifying remarks on his work. The authors thank L. Bombelli for pointing out some typos and spelling errors in the initial version. Having revisited this work, the third author, E. Ertugrul pointed out some errors in the initial version and then joined our team to correct these errors. We believe that this final version is free of previous errors and also has some new results on the chemical potential.

The current revision is dedicated to Prof. Metin G\"urses, on the occasion of his 80th birthday and his new position as an emeritus faculty at Bilkent University. Prof. G\"urses has made significant contributions to the theory of gravitation as well as  to mathematical physics and has been a great source of inspiration for the younger generation.  He had a lasting impact on the theoretical physics community, we wish him many more years to enjoy theoretical physics.

\end{document}